\documentclass[11pt,twoside]{article}

\usepackage{asp2006}
\usepackage{epsf}
\usepackage{psfig}
\usepackage{lscape}
\usepackage{natbib}
\usepackage{graphicx}

\begin{document}
\title{The study about the revised $R_{23}-P$ method for metallicity estimates
based on $\sim$20,000 SDSS galaxies }   
\author{S. Y. Yin$^{1,2}$, Y. C. Liang$^{1}$, B. Zhang$^{1,2}$}   
\affil{$^1$National Astronomical Observatories, Chinese Academy of Sciences,
  20A Datun Road, Chaoyang District, Beijing 100012, China \\
$^2$Department of Physics, Hebei Normal University,
Shijiazhuang 050016, China; email: syyin@bao.ac.cn,ycliang@bao.ac.cn
 }    

\begin{abstract} 
   We select $\sim$20000 metal-rich star-forming galaxies from the 
   SDSS-DR2 to study the $R_{23}-P$ method 
   suggested by Pilyugin et al. for metallicity estimates. 
The oxygen abundances 
derived from their revised $R_{23}-P$ method are $\sim$0.19\,dex lower than 
those derived from the previous one, and $\sim$0.60\,dex lower than the
Bayesian abundances obtained by the MPA/JHU group. 
These abundance discrepancies strongly
correlate with the $P$ parameter and weakly depend on the log($R_{23}$)
parameter. 
\end{abstract}

\section{Introduction}   

  The chemical property of interstellar gas is important to trace 
the chemical evolutionary status of galaxy.  Accurate abundance
measurements for the ionized gas in galaxies require the
determination of the electron temperature ($T_e$), which is
usually able to be measured in metal-poor environments
since the needed auroral line [O~{\sc iii}]$\lambda$4363 can only be detected there.
The common method to estimate the oxygen abundances of metal-rich galaxies
(12+log(O/H)$\geq$8.5) is using empirical strong-line ratios,
  such as $R_{23}$,  $R_{23}-P$,  [N II]/H$\alpha$ etc. (Kobulnicky et al. 1999;
  Liang et al. 2006; Yin et al. 2006, Y06).  
   At present, The $R_{23}-P$ method (Pilyugin et al. 2001, 2005, as 
P01, P05) is widely used since 
the derived (O/H) abundances are consistent with those  
derived from $T_e$ (P05; Y06). 
This is specially interesting for 
metal-rich galaxies. However, the derived abundances from P01 and P05 
show some discrepancy, which has been shown by 
about hundred H~{\sc ii} regions by P05.
Here we select a much larger sample of $\sim$20,000 metal-rich 
star-forming galaxies
from the SDSS-DR2 database to study this effect in more details.
Moreover, we specially compare the abundances derived from the
revised $R_{23}-P$ method given by P05 with 
the Bayesian abundances obtained by Tremonti et al. (2004)
for these SDSS galaxies. 
The selected sample galaxies have 
12+log(O/H$)_{\rm Bay} >$ 8.5 and 12+log(O/H$)_{\rm P05} >$ 8.25, which 
guarantee they are in the upper branch of metallicity.

\section{Comparisons between the abundances  from different methods} 
   
   Previously, P01 gave the $R_{23}-P$ calibration formula for metal-rich 
   galaxies on the basis of $\sim$40 H~{\sc ii} regions. P05 revised
   this calibration by using an extended sample, about 104 H~{\sc ii} regions.
   P05 has shown the difference between these two calibrations.
   Here we
   use a much larger sample to study this discrepancy.
  Figs.1a-c show the large discrepancy between the two derived abundances
  from P01's and P05's calibrations, 
  and the relations of this discrepancy with 
  log($R_{23}$) and $P$. Then, we also study the large discrepancies between
  the
  Bayesian abundances and those derived from P05's calibration, and their relations with 
  log($R_{23}$) and $P$ (Figs.1d-f). See Caption of Fig.1 for more details.
  
\begin{figure}
\centering
\psfig{file=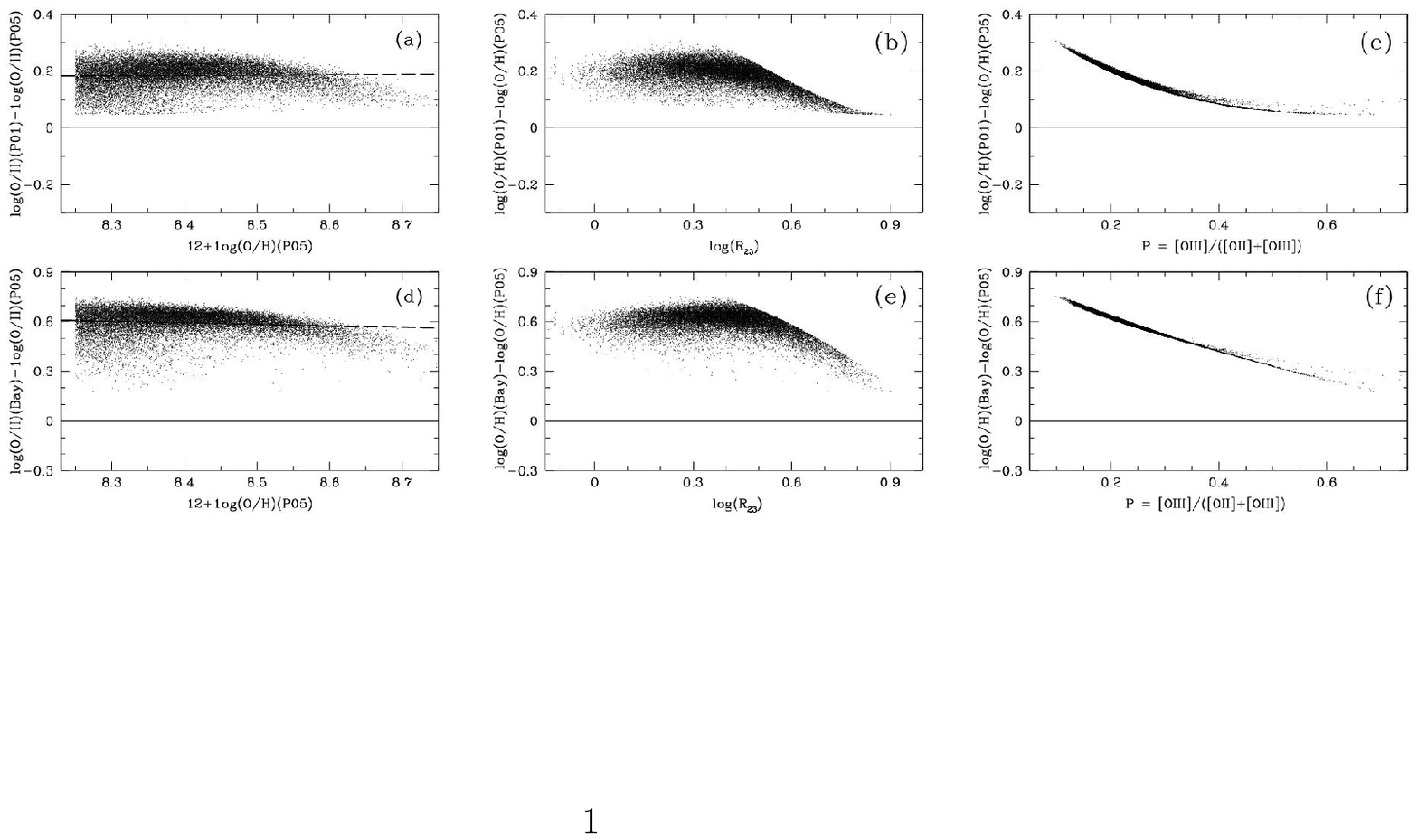,width=11.8cm,angle=0,%
    bbllx=116pt,bblly=199pt,bburx=575pt,bbury=365pt,clip=} 
\caption {Comparisons between the oxygen abundances derived from
P01, P05 and Bayesian technique.
{\bf (a)} The differences between the log(O/H) abundances derived
from the previous (P01) and revised (P05)
$R_{23}-P$ methods 
as a function of the oxygen abundances derived from the revised $R_{23}-P$.
 The differences reach $\sim$0.19\,dex.
{\bf (b)} The differences between the log(O/H) derived from P01 and P05
as a function of log($R_{23}$).
It is  constant ( $\sim$0.2\,dex) as log($R_{23}$) 
$<$ 0.5, but becomes smaller following the increase of the log($R_{23}$), 
which may be due to the turn-over region of the $R_{23}$ parameter.
{\bf (c)} The abundance differences 
as a function of $P$ parameter. It shows a tight correlation with $P$, 
specially for the sample with smaller value of $P$ 
($P$$<$ 0.5), but becomes constant as $P$ $>$ 0.5.
{\bf (d)},{\bf (e)},{\bf (f)} show the oxygen abundance 
differences between the Bayesian abundances and 
those derived from the revised $R_{23}-P$ method of P05 
as a function of log(O/H)$_{\rm P05}$,
the log($R_{23}$), and $P$ parameters, respectively.
{\bf Fig.(d)} shows the $\sim$ 0.60 dex difference.
{\bf Fig.(e)} shows similar results as {\bf Fig.(b)}.
{\bf Fig.(f)} shows that the oxygen abundance differences 
have a steep linear correlation with the $P$ parameter.
}
\end{figure} 
 
  In summary, different metallicity calibrations can result in very different
  abundances, and we should be careful when compare the abundance estimates
  derived from different methods.
  
\acknowledgements 
     
      This work was supported by the National Science Foundation of China 
(NSFC) Foundation under No.10403006, 10373005, 10433010, 10573022.


\begin{thebibliography}{}

\bibitem[1999] {K99} Kobulnicky, H.A., Kennicutt, R.C.Jr., \& Pizagno, J.L., 
1999, ApJ, 514,544
\bibitem[2006] {L06} Liang, Y.C., Yin, S.Y., Hammer, F. et al. 2006, ApJ, 652, 257 
\bibitem[2001] {P01} Pilyugin, L.S., 2001, A\&A, 369, 594 (P01)
\bibitem[2005] {P05} Pilyugin, L.S., \& Thuan, T.X, 2005, ApJ,631, 231 (P05)
\bibitem[2004] {T04} Tremonti, C.A., Heckman, T.M., Kauffmann, G., et al. 
2004, ApJ, 613, 898
\bibitem[2006] {Y06} Yin, S.Y., Liang, Y.C. et al., 2006, A\&A (in press), astro-ph/0610068 (Y06)
\end{thebibliography}
\end{document}